\documentclass[referee]{ctextemp_raa}            

\usepackage{ctextemp_graphicx,times}             
\usepackage{ctextemp_natbib}
\usepackage{amssymb,amsmath}
\usepackage{color}

\bibpunct{(}{)}{;}{a}{}{,}

\usepackage[a4paper=true,dvipdfm=true,pagebackref=true]{hyperref}
\hypersetup{colorlinks = true, linkcolor = green, anchorcolor = red, citecolor = blue, filecolor = red, pagecolor = red, urlcolor = red}

\begin{document}

\title{Energy and spectral analysis of confined solar flares from radio and X-ray observations$^*$
\footnotetext{\small $*$ Supported by the National Natural Science Foundation of China.}}

\volnopage{ {\bf 20XX} Vol.\ {\bf X} No. {\bf XX}, 000--000}
\setcounter{page}{1}

\author{Chengming Tan\inst{1,2}, Karl-Ludwig Klein\inst{3}, Yihua Yan\inst{1,2}, Satoshi Masuda\inst{4}, Baolin Tan\inst{1,2}, Jing Huang\inst{1,2}, Guowu Yuan\inst{5}}

\institute{ CAS Key Laboratory of Solar Activity, National Astronomical Observatories of Chinese Academy of Sciences, Datun Road A20, Chaoyang District, Beijing 100101, China; {\it tanchm@nao.cas.cn}\\
\and
School of Astronomy and Space Sciences, University of CAS, Beijing 100049, China\\
\and
Observatoire de Paris, LESIA \& Station de radioastronomie de Nan\c{c}ay, Univ. PSL, CNRS, Sorbonne Univ., Univ. de Paris, Univ. d'Orl\'eans, 5 place Jules Janssen, F- 92190 Meudon, France \\
\and
Institute for Space-Earth Environmental Research, Nagoya University, Nagoya 464-8601, Japan\\
\and
School of Information Science and Engineering, Yunnan University, Kunming 650091, China\\
\vs \no
{\small Received 20XX Month Day; accepted 20XX Month Day}}

\abstract{The energy and spectral shape of radio bursts may help us understand the generation mechanism of solar eruptions, including solar flares, CMEs, eruptive filaments, and various scales of jets. The different kinds of flares may have different characteristics of energy and spectral distribution. In this work, we selected 10 mostly
confined flare events during October 2014 to investigate their overall spectral behaviour and the energy emitted in microwaves by using radio observations from microwaves to interplanetary radio waves, and X-ray observations of GOES, RHESSI, and Fermi/GBM. We found that: All the confined flare events were associated with a microwave continuum burst extending to frequencies of $9.4\sim15.4$ GHz, and the peak frequencies of all confined flare events are higher than 4.995 GHz and lower than or equal to 17 GHz. The median value is around 9 GHz. The microwave burst energy (or fluence) as well as the peak frequency are found to provide useful criteria to estimate the power of solar flares. The observations imply that the magnetic field in confined flares tends to be stronger than that in 412 flares studied by \citealt{Nita+2004}. All 10 events studied did not produce detectable hard X-rays with energies above $\sim$300 keV indicating the lack of efficient acceleration of electrons to high energies in the confined flares.
}

\authorrunning{C.M. Tan, K.-L. Klein, Y.H. Yan, et al. }            
\titlerunning{Energy and spectral analysis of the confined solar flares from radio and X-ray observations}  
\maketitle

%
\section{Introduction}           
\label{sect:intro}
Solar flares are believed to be the most powerful explosions in the solar system. They may produce strong impact on the space weather around the Earth. However, the flares are characterized by diversity and complexity. For example, the energy released in a flare may range from above $10^{26} J$ in a super flare ($>$X10 flare) to less than $10^{21} J$ in an A-class small flare. As for their timescale, the duration may range from a few seconds to above 2 hours. So far, there is nearly a consensus that the flare energy is released through magnetic reconnection in the magnetized plasmas of the solar corona. The possible mechanism that trigger magnetic reconnection and the flare include the emerging magnetic flux model (\citealt{Heyv+1977}), the standard flare model (\citealt{Shib+1995}), tether-cutting model (\citealt{Moore+2001}, circular ribbon flare model (\citealt{Wang+Liu+2012}), and magnetic-gradient driven model (\citealt{Tanb+2020}), etc. Many flares are accompanied with very spectacular coronal mass ejections (CME). They are usually called eruptive flares, while at the same time there are also many flares without accompanying CME, which are called confined flares (\citealt{Ji+2003}).

With such strong diversity of solar flares, it is necessary to investigate the characteristics in different flare populations, respectively. Such as in different class flares, flares in different phases of solar cycles (valley, ascend, peak and descend phases of solar cycles, defined as \citealt{Tanb+2019}), long-duration flares or short-duration flares, eruptive flares or confined flares, etc.

One of the main characteristics of solar flares is their energy and spectral distribution which may help us to understand the generation mechanism of solar eruptions, including solar flares, CMEs, eruptive filaments, and various scales of jets (\citealt{Asch+1998}), and to obtain the evolution of solar activity. Many people investigated the microwave spectrum of solar flares and its correlation with hard X-rays, soft X-rays, optical, radio wavelengths, and energetic particles in interplanetary space (\citealt{Guid+Cast+1975, Asch+1998, Isli+Benz+2001, Trot+2015}, etc.). \citealt{Nita+2002, Nita+2004} investigated the peak flux distribution of solar radio burst data observed during 1960-1999 as a function of frequency and time over a wide frequency range, and found that the distributions are well fitted by power laws over a wide range of flux densities, but diverge from a power law at both high and low flux densities. They also found that the divergence of the distribution from the fit at low flux density is probably due to sensitivity limits of the recording instruments which ranges from about 10 sfu during solar minimum to about 20 sfu during solar maximum. \citealt{Furst+1971} and \citealt{Meln+2008} studied the changes in spectral peak frequency of microwave bursts with time and found that, for the majority of simple microwave bursts, the peak frequency shows a high positive correlation with flux density, which is in qualitative agreement with theoretical expectations based on gyrosynchrotron self-absorption. \citealt{Kruc+2020} found that there is a highly significant correlation between the microwave (17 GHz and 34 GHz) peak flux and the nonthermal hard X-ray bremsstrahlung peak emission seen above 50 keV. However, the existing published studies do not distinguish different types of flares, and we believe that different kinds of flares may have different energy and spectral characteristics.

In the present paper we focus on confined flare events to study their energy and spectral distributions, including the radio peak flux, the fluence or energy emitted at microwave frequencies and in soft and hard X-rays, as well as the relationships among them. A series of confined flares from one active region in October 2014 is a particularly appropriate opportunity to study flares and electron acceleration under conditions where the ambient magnetic field does not change fundamentally between individual events, and to search for correlations between different manifestations of energy release. Section 2 introduces the related observational data and definitions and extractions of parameters. Section 3 presents the main statistical results. The conclusions and the related discussions are summarized in Section 4.


\section{Event survey and data analysis}
\label{sect:data}

In this work, we select 10 confined flares to investigate their spectral evolution from microwave to interplanetary radio wavelengths, and X-ray observations of GOES, Fermi/GBM, and RHESSI. Theses events occurred in solar active region AR12192 during October 2014. They were investigated by \citealt{Chen+2015} using the observations from the Atmospheric Imaging Assembly (AIA) and Helioseismic and Magnetic Imager (HMI) on board the Solar Dynamics Observatory (SDO).

The data used includes the radio and X-ray observations. Fixed frequency observations are provided by Nobeyama Radio Polarimeters (NoRP; \citealt{Naka+1985}) at 1.0, 2.0, 3.75, 9.4, 17, and 35 GHz, and by the Radio Solar Telescope Network (RSTN) at 0.245, 0.41, 0.61, 1.415, 2.695, 4.995, 8.80, 15.4 GHz. The RSTN monitors 8 fixed frequencies at four observatories over the world. They are located at Sagamore Hill (Massachusetts, U.S.A), Palehua (Hawaii, U.S.A), Learmonth (Australia), and San Vito (Italy). The spectral observations included the Chinese Solar Broadband Radio Spectrometer at Huairou (SBRS/Huairou: \citealt{Fu+1995, Fu+2004}) at 1.1-7.6 GHz, Mingantu Spectral Radioheliograph (MUSER: \citealt{Yan+2016, Yan+2021}) at 2-15 GHz, decimetric and metric spectrometers of the Yunnan Astronomical Observatories (YNAO: \citealt{Gao+2014}) at 70-700 MHz, the Nan\c{c}ay Decameter Array (NDA,\citealt{Leca+2000}) located at Nan\c{c}ay, France at 20-70 MHz, the ORFEES spectrograph (located at Nancay, France) at 144-1000 MHz, the Culgoora spectrograph (Australia) at 18-1800 MHz, and the Wind/WAVES radio spectrograph (\citealt{Boug+1995}) at 20 kHz-13.825 MHz. The X-ray observation includes the soft X-ray (SXR) flux at 1-8{\AA} and 0.5-4{\AA} observed by GOES (\citealt{Born+1996}), \and hard X-ray (HXR) flux observed by the Fermi Gamma-ray Burst Monitor (Fermi/GBM, \citealt{Meegan+2009}) and RHESSI (\citealt{Lin+2002}).

\begin{table}[ht]
\bc
\begin{minipage}[]{150mm}
\caption[]{The main information of GOES soft X-ray and radio emission of the selected events\label{tab1}}\end{minipage}
\setlength{\tabcolsep}{1pt}
\small
 \begin{tabular}{cccccccccccc}
  \hline\noalign{\smallskip}
  1  &  2  &  3  &  4  &  5 &  6 &   7    &      8        &       9        &     10    &     11     &     12    \\
ID   &Class& Date&Start&Max.&End &IP burst&               &SBRS observation&           & NORP, RSTN & Peak freq.\\
     &     &     &     &    &    &        &   1.1-2.1GHz  &   2.6-3.8GHz   & 5.2-7.6GHz& burst GHz  &       GHz \\
  \hline\noalign{\smallskip}
MF1* &M1.6 &Oct18&0702 &0758&0849&   no   &    dm.FS      &     cont.      &  cont.    &0.245-15.4  &     4.995\\
MF8* &M8.7 &Oct22&0116 &0159&0228&   no   &dm.FS,and cont.&     cont.      &  cont.    &0.245-35.0  &       9.4\\
MF9  &M2.7 &Oct22&0511 &0517&0521&   no   &       no      &        no      &  cont.    &4.995-35.0  &      17.0\\
MF11*&M4.0 &Oct24&0737 &0748&0753&type III&dm.FS,and cont.&     cont.      &  cont.    &0.245-15.4  &       8.8\\
MF16 &M7.1 &Oct27&0006 &0034&0044&   no   &       no      &        no      &  cont.    &4.995-35.0  &      17.0\\
MF17 &M1.0 &Oct27&0144 &0202&0211&   no   &       no      &        no      &  cont.    &4.995-17.0  &      17.0\\
MF18*&M1.3 &Oct27&0335 &0341&0348&   no   &       no      & cm.FS          &  cont.    & 2.84-35.0  &       9.4\\
MF21*&M3.4 &Oct28&0215 &0242&0308&   no   &    dm.FS      &      no data   &  cont.    &  2.0-35.0  &      17.0\\
MF22 &M6.6 &Oct28&0323 &0332&0341&   no   &     bad data  &    bad data    &  bad data &  9.4-35.0  &      17.0\\
MF24 &M1.0 &Oct29&0603 &0820&0852&   no   &       no      &   weak cont.   &  cont.    &2.695-15.4  &      15.4\\
  \noalign{\smallskip}\hline
\end{tabular}
\ec
\tablecomments{1.0\textwidth}{The 1st column is the event ID of GOES SXR M-class flare, same as \citealt{Chen+2015}. The star * markes the event with dm. or cm. fine structure. The 2nd column is the flare class. The 3rd column is the event date. The 4-6th columns are the begin, peak, and end time of the SXR flare. The 7th column indicates the interplanetary (IP) burst at decametre-hectometre wavelengths (DH). The 8-10th indicates the fine structure bursts or continuum bursts observed by SBRS/Huairou at three bands, respectively. The 11th column indicates the burst at the fixed frequency observed by NORP or RSTN. The last column lists the peak flux frequency that analyzed from NoRP or RSTN observations. dm.FS or cm. FS indicates the spectral fine structure at decimeter or centimeter wavelength. cont. indicates the continuum enhancement.}
\end{table}

Table.1 lists the main information of GOES SXR and radio emission of the selected ten events. Columns 4-6 are the start, peak, and end time of the SXR flare. Column 7 indicates whether an interplanetary type III burst was observed. Columns 8-10 give a brief characterization of different frequency ranges as to the presence of
decimeter (abbreviation as dm.) or centimeter (abbreviation as cm.) fine structure (abbreviation as FS), and microwave continuum (abbreviation as cont.). Section 3 will illustrate in detail the relationship between the spectral characteristics and the flux density spectrum. Columns 11 and 12 indicate the range and peak frequency of the fixed-frequency observations. Nine events were associated with microwave continuum burst at frequencies greater than 5 GHz, the MF22 event was associated with a microwave continuum burst at frequency of greater than 9.4 GHz.
The peak frequencies of all events are above 4.995 GHz.

Figure.1 presents the dynamic spectra of the ten confined flare events, as observed by Wind/Waves, Culgoora, NDA, Ynao, ORFEES, and SBRS/Huairou. Five events are accompanied by groups of dm. or cm. fine structure, which are presented in Figure.2.
\begin{figure}[ht]
   \centering
   \includegraphics[width=14.5cm, angle=0]{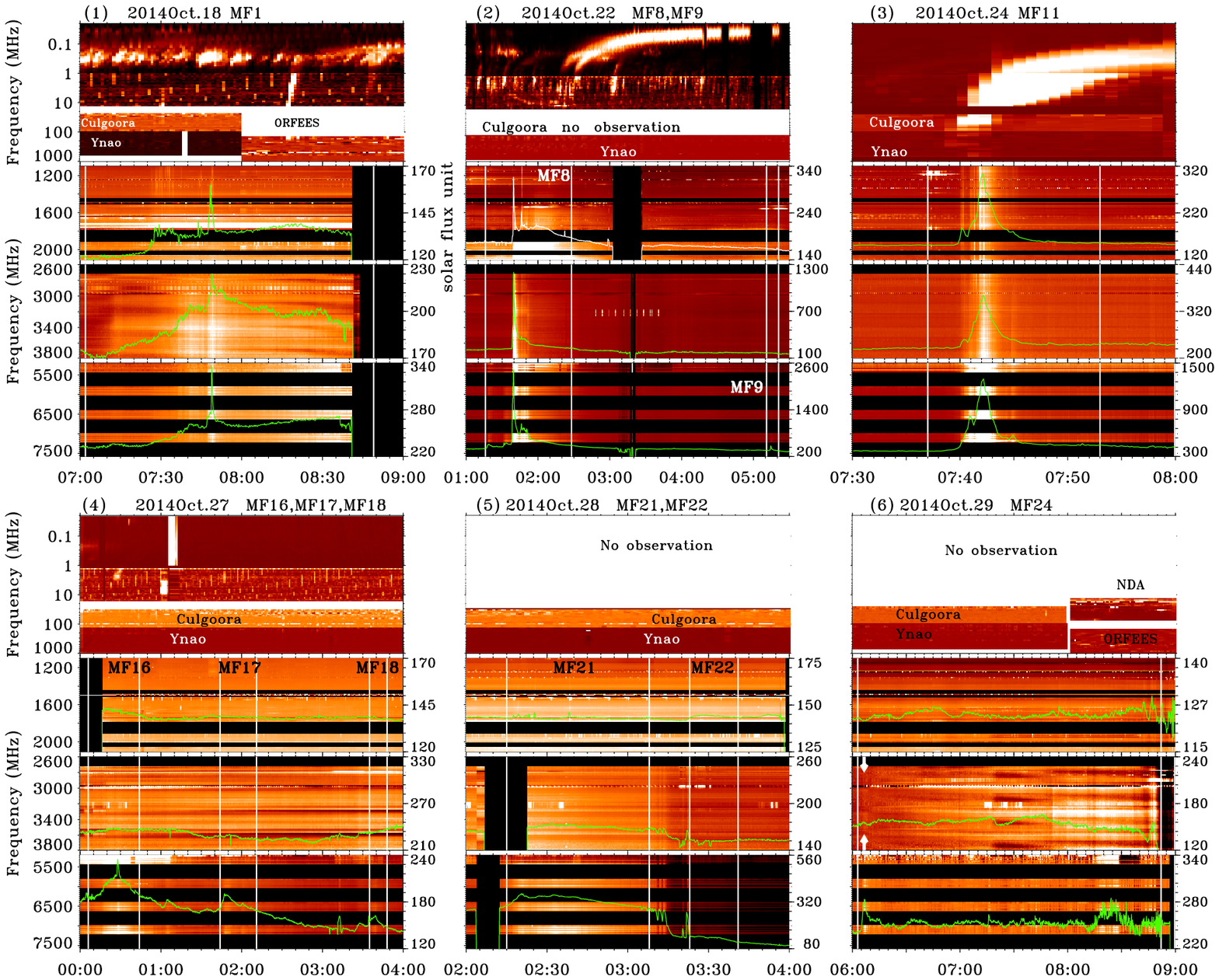}
   \caption{ Radio spectra of ten flare events on six days of October 2014. Each panel presents a daily spectrum obtained from Wind/Waves, Culgoora, NDA, Ynao, ORFEES, and SBRS/Huairou. The white lines mark the begin and end time of the event. The green curves are the flux profiles at 1356MHz, 2840MHz, and 5900MHz, respectively. The tick values at the right side of the panel indicate the flux density in solar flux units. The 5.2-7.6 GHz band is only partially observed.}
   \label{Fig1}
   \end{figure}

\begin{figure}[ht]
   \centering
  \includegraphics[width=14.5cm, angle=0]{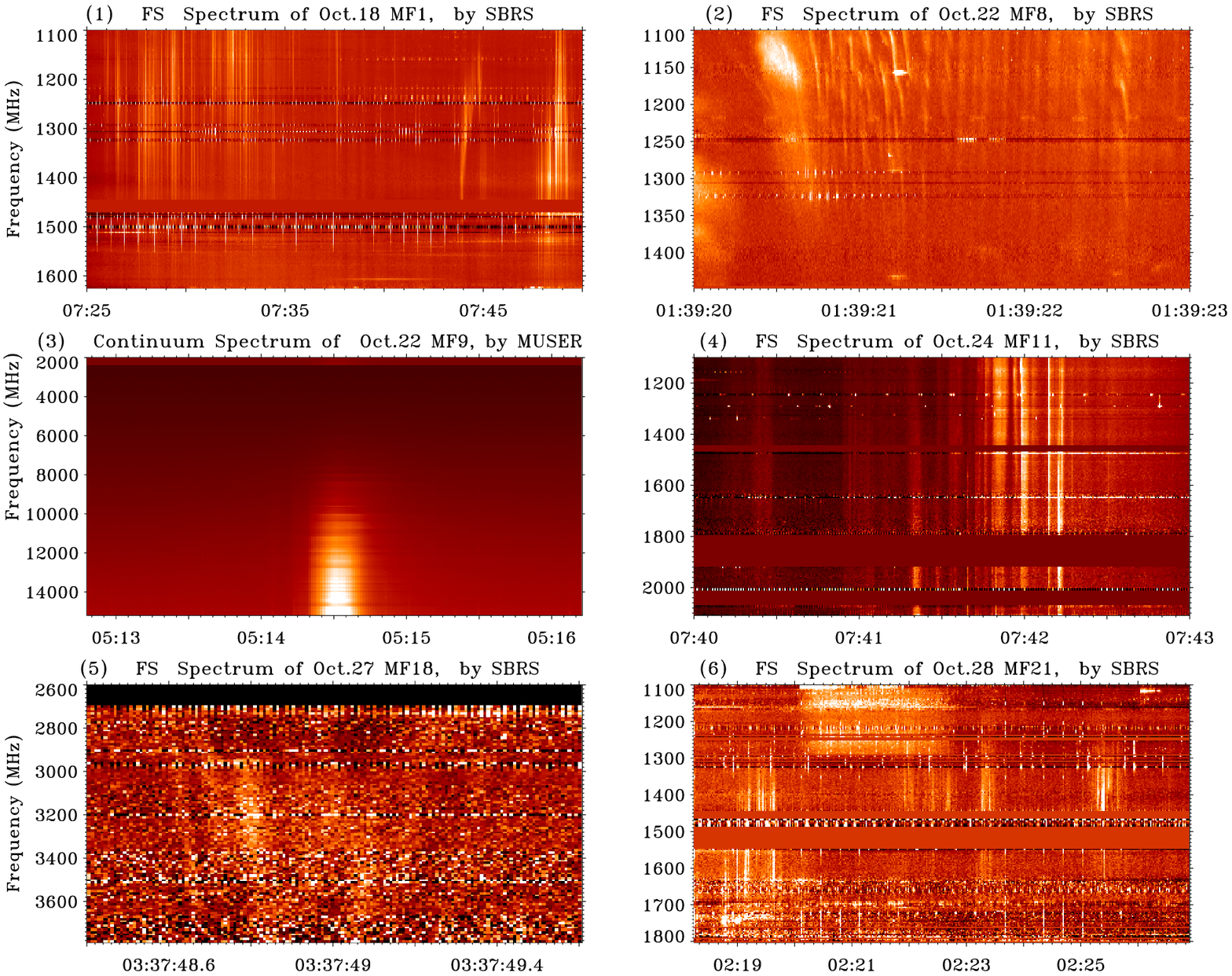}
   \caption{The spectrum of fine structures (abbreviation as FS) in five events observed by SBRS and one impulsive microwave continuum burst observed by MUSER.}
   \label{Fig2}
   \end{figure}

Panel (1) of Figure 1 presents the observed spectrum on 2014 Oct.18 (event MF1, 07:02-08:49 UT). The start and end time of MF1 are marked by white lines. The green curves are flux profiles at single frequencies of 1356 MHz, 2840 MHz, and 5900 MHz, respectively. The event shows no spectral fine structure on the microwave continuum burst in the frequency range of 2.6-3.8GHz and 5.2-7.6 GHz. There are spectral fine structures on the continuum burst of 1.1-2.1GHz. Panel (1) of Figure.2 presents the details of the fine structures. They are groups of type III bursts with fast frequency drift rates. At lower frequencies the Culgoora and Wind/Waves radio spectra (20 kHz$\sim$100 MHz) showed no counterpart of event MF1.

Panel (2) of Figure 1 presents the observed spectrum on 2014 Oct.22 (events MF8, 01:16-02:28 UT, and MF9, during 05:11-05:21UT) observed by Wind/Waves, Ynao, and SBRS/Huairou. There are continuum bursts in the frequency range of 2.6-3.8 GHz and 5.2-7.6 GHz around the MF8 event, without spectral fine structure. Groups of type III bursts with fast downward and upward drifts are seen on top of the continuum in the frequency range 1.1-2.1GHz. Panel (2) of Figure.2 presents the details of this fine structures. The Ynao radio spectrum showed no counterpart to this event. The Wind/Waves spectrum showed an IP type III burst about 30 minutes afterwards, which had no connection with the main microwave burst. Around the MF9 event, there is only an impulsive burst in the frequency of 5.2-7.6 GHz observed by SBRS/Huairou and in the frequency range $4.0\sim15$ GHz observed by MUSER without spectral fine structure. This is presented in panel (3) of Figure.2.

Panel (3) of Figure 1 presents the observed spectrum on 2014 Oct.24 (event MF11, 07:37-07:53UT) observed by Wind/Waves, Culgoora, Ynao, ORFEES, and SBRS/Huairou. All bands of SBRS/Huairou observed the continuum burst. There are spectral QPP structures in the frequency range of 1.1-2.1 GHz, which are presented in panel (4) of Figure.2. At low frequencies, the Ynao, Culgoora radio, and Wind/Wave spectrum at 25-700MHz showed a typical type III burst. The presence of type III emission at meter-to-kilometer wavelengths is an exceptional finding in confined flares in general (\citealt{Gopa+2009,Klein+2010}), and in the flares from the October 2014 active region in particular. \citealt{Chen+2015} reported that this event originated from the periphery of the active region, and was accompanied by a jet. We concluded that it is because of this exceptional location that the electrons had access to open field lines extending to the high corona and into the interplanetary space.

Panel (4) of Figure 1 presents the observed spectrum on 2014 Oct.27 (event MF16, 00:06-00:44 UT, MF17, 01:44-02:11UT, and MF18, 03:35-03:48UT) observed by Wind/Waves, Culgoora, Ynao, and SBRS/Huairou. During these three events, there are continuum bursts in the frequency of 5.2-7.6 GHz, while no continuum burst is observed in the ranges 2.6-3.8 GHz and 1.1-2.1 GHz. Around the MF18 event in the frequency of 2.6-3.8 GHz, there is a small group of very weak ($S_{burst}\approx S_{bg}+3\sigma$) and very short duration ($<1$ second) of  drifting burst which presented in panel (5) of Figure.2. Below 1.0 GHz there is no counterpart of any of these events.

Panel (5) of Figure 1 presents the observed spectrum on 2014 Oct.28 (events MF21, 02:15-03:08UT, and MF22, 03:23-03:41UT) observed by Culgoora, Ynao, and SBRS/Huairou. For the MF21 event, a continuum burst is observed in the frequency range of 5.2-7.6 GHz and several groups of narrowband weak fine structures with fast drifts in the frequency range of 1.1-2.1 GHz. They are presented in panel (6) of Figure.2. For the MF22 event, there is no radio burst at any band below 7.6 GHz.  Panel (5) of Figure.3 shows only radio emission at frequencies  above 9.4 GHz.

Panel (6) of Figure 1 presents the observed spectrum on 2014 Oct.29 (event MF24, 06:03-08:52UT) observed by Culgoora, Ynao, ORFEES and SBRS/Huairou. There are continuum bursts at the beginning (06:05UT) and peak time (08:20UT) of the event, in the frequency range of 5.2-7.6 GHz. There is only a very weak continuum burst (indicated by two white arrows) at the beginning (06:05UT) of the event in the frequency of 2.6-3.8 GHz.

\begin{figure}[ht]
\centering
\includegraphics[width=14.5cm, angle=0]{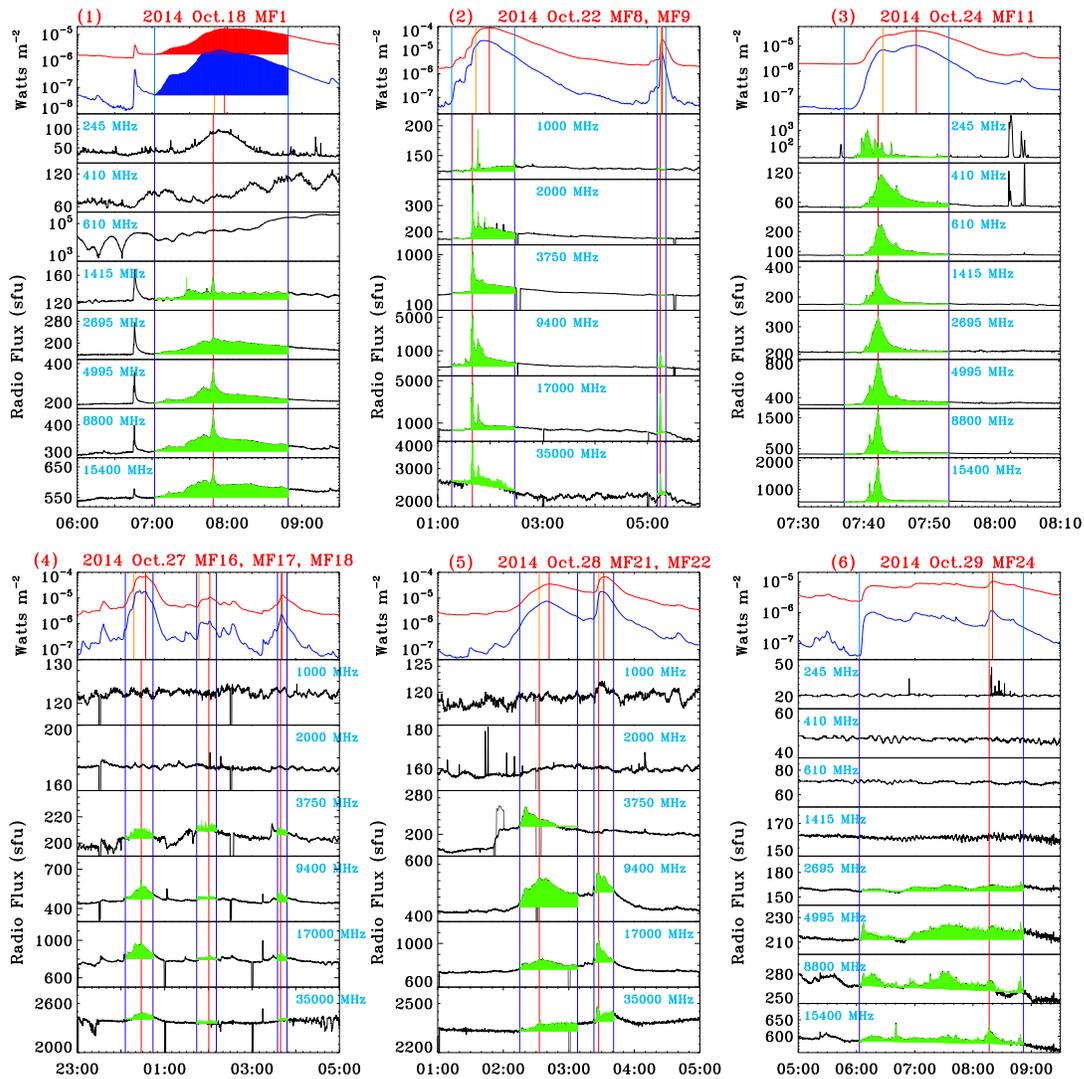}
\caption{Composed figure of GOES SXR flux and radio flux of ten events during six days. For each day, the GOES-15 soft X-ray flux is plotted in the top mini panel, and the radio flux of RSTN or NORP observation at several frequencies in the bottom mini panels. As an example, top of panel (1), the integrated SXR flux of the flare after background subtraction is displayed by red (1-8{\AA}) and blue (0.5-4{\AA}) shading. The vertical blue and red lines mark the start-end time and the peak time, respectively. The vertical orange line is the HXR peak time. The green area is the integrated radio flux of the burst after background subtraction.}
\label{Fig3}
\end{figure}

Figure 3 presents the composed figure of GOES SXR flux and radio flux of the ten confined flare events. For each day's panel, the top mini panel shows the profiles of GOES SXR flux at 1-8{\AA} (red curve) and 0.5-4{\AA} (blue curve). The start-end time and peak time are marked as vertical blue line and red line, respectively. In the bottom mini panels the radio flux is plotted at eight frequencies (245, 410, 610, 1415, 2695, 4995, 8800, and 15400 MHz) observed by Palehua (Hawaii) of RSTN or at six frequencies (1000, 2000, 3750, 9375, 17000 and 35000 MHz) observed by NORP. The start-end time (blue line) are the same as SXR event time. The peak time (red line) are calculated as maximum time of microwave burst considering most frequencies. For the long duration event with multiple peak time, for example the event of 2014 Oct.29, only the closest one to the SXR peak time was selected. The peak time of radio flux are about 1$\sim$20 minutes earlier than that of SXR. This is consistent with the result of \citealt{Kruc+2020}. Figure 3 also shows that the rapid increase of SXR light curve is very close to the maximum of HXR flux (orange line) or the microwave light curves. This is consistent with the deduction of the Neupert effect (\citealt{Neupert+1968}, \citealt{Ning+Cao+2010}). The green areas designated the integrated radio flux in events and at frequencies where the burst was clearly identified. In the panel of 2014 Oct.18, the integrated radio flux (green area) of the MF1 event at 245, 410, and 610 MHz was not plotted because of strong radio interference. In the panel of 2014 Oct.27, the time range (blue line) of the third event (MF18) did not include the radio burst just before the event. But the flux value of this event is still kept  for the later energy analysis. In the panel of 2014 Oct.28, the mini panel of radio observations shows the flux profile had some jumps (gray line) around the burst. The jumps were corrected as shown by the black lines. The MF22 event showed the radio burst happened only above 9.4 GHz. In the panel of 2014 Oct.29, the MF24 is a long duration event with multiple flares. The radio flux of the background decreased or increased along the time because of some uncertain reason. A time-varying background had to be subtracted to calculate the flux of the burst. The calculation of background and subtraction were indicated in the quantitative analysis part after.

\begin{figure}[h]
\centering
\includegraphics[width=14.5cm]{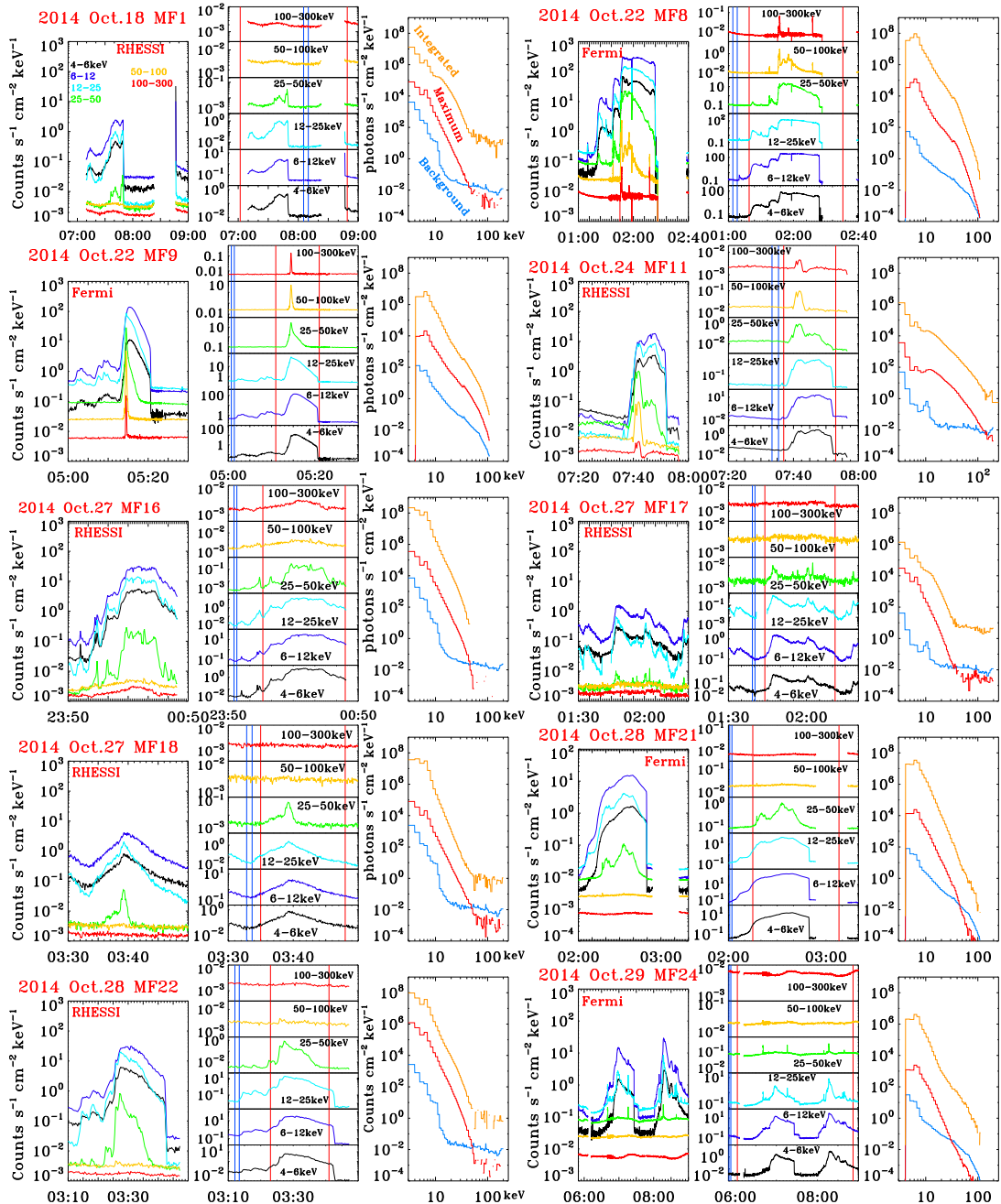}
\caption{\small {Fermi or RHESSI HXR counts flux and photons flux spectra of ten events. For each event, the HXR flux of 6 energy bands are plotted in the left mini panel by different colors. The vertical red and blue lines mark the event time and the background intervals, respectively. In the panel of the photons flux spectrum the blue lines show the flux spectra of the background. The red lines show the peak flux spectra after background subtraction. The orange lines are the integrated photons flux (fluence) spectra after background subtraction.}}
\label{Fig4}
\end{figure}

The Fermi or RHESSI HXR counts flux profiles of the ten confined flare events are shown in Figure.4. Some events are with good Fermi observation, while some events are not. Only the better one was included. Most of the events have no credible flux enhancement above $\sim50$ keV. All the events has no credible flux enhancement at energy of higher than 300 keV. For each event, the right panel shows the photons flux spectra of the event. Both the counts flux and photons flux spectrum were analyzed with solar software (SSW). The SSW supplied sets of program and tools. Thus the counts flux and photons flux can be obtained. The integrated photon flux (fluence) spectrum is the integration of all the photon flux within the event time (red lines in figure.4).

In the following we conduct a quantitative analysis in order to search for correlations. For microwave bursts we consider the peak flux density, the flux density integrated over the entire event at a given frequency (fluence), and the total energy radiated during the microwave burst. For reference we consider the SXR peak flux and fluence.

(1) Background radio flux ($S_{bg}(\nu)$): The background flux density, which must be subtracted, was calculated as the average value during a time interval lasting several minutes outside the GOES SXR event, avoiding any period of enhanced solar radio emission or terrestrial interference. The standard deviation was also calculated during this interval, and the time of the burst was considered as the time interval where the flux density exceeds the background plus three times the standard deviation. The solar flux unit is $1sfu=10^{-26}J\cdot s^{-1}\cdot m^{-2}\cdot Hz^{-1}$.

(2) Peak radio flux ($S_{peak}(\nu)$): This is the peak flux value during the event (vertical red line in Figure.3), after subtraction of the pre-event background. In the microwave band $>1$ GHz, the peak times are almost the same at different frequencies in most cases.

(3) Radio fluence at a single frequency ($Fluen(\nu)$): This is the integral of the radio flux density during the burst after background-subtraction (the green area in Figure.3) at a single frequency $\nu$. For some events the $Fluen(\nu)$ at low frequencies were not plotted because of interference. Practically $Fluen(\nu)$ is calculated as the sum
\begin{equation}
Fluen(\nu)={\sum_{i=m}^n(\frac{1}{2}[S_{i+1}+S_{i}-2\times S_{bg}]\times[t_{i+1}-t_{i}])}
\label{eq:lab1}
\end{equation}
Here, $S_{i}$ is the instantaneous radio flux of the burst, $\nu$ is frequency, $i=m\rightarrow n$ is the data number, $t$ is the time in seconds. Equation (1) gives the fluence of the radio burst in units of $J\cdot m^{-2}\cdot Hz^{-1}$.

(4) Radio burst total energy ($E_{total}$): The total energy of the radio burst can be calculated by summing the fluences over frequency, as in equation (2), using only interference-free frequencies above 1 GHz. The unit is $J\cdot m^{-2}$.
\begin{equation}
E_{total}={\sum_{i=m}^n(\frac{1}{2}[Fluen(\nu_{i+1})+Fluen(\nu_{i})]\times[\nu_{i+1}-\nu_{i}])}
\label{eq:lab2}
\end{equation}

(5) Background SXR flux ($F_{SXR\underline{~}bg}$): It was calculated as the average value of several minutes before the GOES SXR event time (vertical blue line in Figure.3). The SXR flux unit is $Watt\cdot m^{-2}$.

(6) Peak SXR flux ($F_{SXR\underline{~}peak}$):  This is the peak value during the SXR flare event. Usually it is defined as the SXR flare class .

(7) SXR fluence ($Fluen_{SXR}$): It was calculated as the integral of the background-subtracted SXR flux during the flare at a single band. The unit is $J\cdot m^{-2}$. In the top panel (1) of figure.3, the red area is $Fluen_{SXR}$ at 1-8{\AA} (A band) and the blue area is $Fluen_{SXR}$ at 0.5-4{\AA} (B band). It can be calculated with the analogue of equation (1).

In the next section, we will present the statistical results of the energy, spectral characteristics and the related correlations.

\section{Result}
\label{sect:result}
This section illustrates the results of our analysis. The first subsection analyzes the spectral characteristics of the ten confined flare events, the relationship between the spectral characteristics and the flux spectrum, and the relationship between the spectral characteristics and the fluence spectrum. The spectral characteristics include the microwave continuum, centimeter and decimeter spectral fine structures, and bursts at meter-kilometer wavelengths. The second subsection analyzes the relationships between microwave burst and GOES SXR flare. The third subsection analyzes the relationships between microwave burst and HXR flare.

\begin{figure}[ht]
   \centering
  \includegraphics[width=12cm, angle=0]{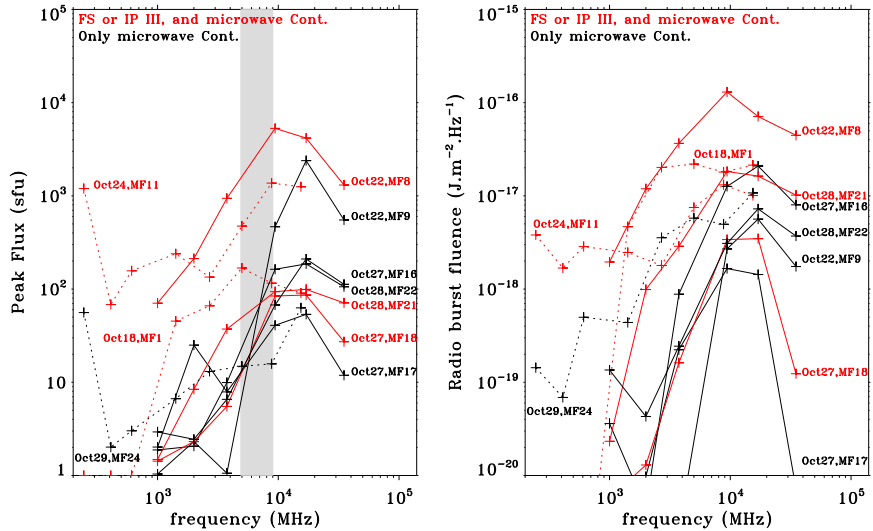}
   \caption{The flux and fluence spectral characteristics of the confined flares. The left panel presents the peak flux spectrum. The grey shading is the median range ($4.9\sim9$ GHz) of peak frequency reported by \citealt{Nita+2004}. Right panel plotted the radio burst fluence versus frequencies. The flux value is calculated after subtracting the background radio flux.}
   \label{Fig5}
   \end{figure}

\subsection{The spectral characteristic of the confined flares}

According to the above analysis in Section 2, we may classify ten confined flare events into 2 classes. The first class events were accompanied by decimeter or centimeter spectral fine structure or IP burst (only one case), and also microwave continuum burst. The microwave spectra of this class are plotted in red color in Figure 5. The 2nd class events have only microwave continuum at frequencies higher than 2.695 GHz, without spectral fine structure or IP burst. The spectra are plotted in black color in Figure 5.

Figure 5 displays the peak flux spectra and the fluence spectra of ten confined flare events. The left panel shows the peak flux spectrum. The flux value is calculated after subtracting the background radio flux $S_{bg}(\nu)$. The calculation was illustrated in section.2 and Figure 3. In the microwave band, Table 1 and the left panel of Figure.5 show that the peak frequency of all events is greater than or equal to 4.995 GHz. The peak frequency of the events with fine structures is lower than or equal to 17 GHz. The peak frequency of the events without fine structures is $15.4\sim17$ GHz. In the right panel the radio burst fluence is plotted versus frequency. The radio burst fluence of the events with fine structure tends to be higher than that of the events without fine structure. The event of 2014 Oct. 29 MF24 is an unusual long duration event with multiple energy releases, thus the radio burst fluence is a little higher than other events without fine structure. The MF18 burst has the lowest fluence among the events with fine structure, but the fine structure itself is also faint ($S_{burst}\approx S_{bg}+3\sigma$) and of very short duration ($<1$ second). So this is consistent with an overall trend that events with fine structure have higher peak flux and higher fluence than events without fine structure.

\subsection{The relationships between microwave peak flux, fluence, energy and SXR peak flux, fluence}

\begin{figure}[ht]
   \centering
  \includegraphics[width=14.5cm, angle=0]{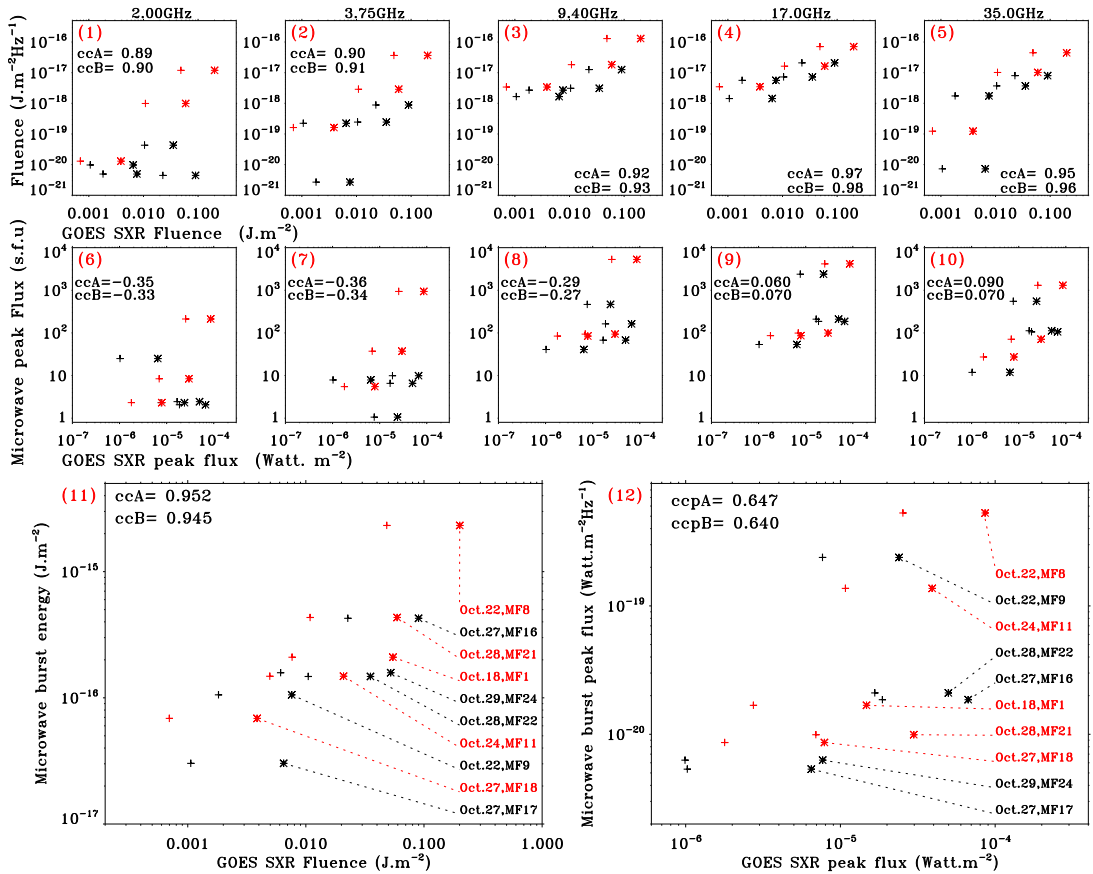}
     \caption{Panel (1)-(5) plotted the microwave burst fluence at five NORP frequencies versus SXR fluence in the two GOES bands. Panel (6)-(10) plotted the microwave burst peak flux at five NORP frequencies versus SXR peak flux in the two GOES bands. Panel (11) plotted the microwave burst energy versus SXR fluence in the two GOES bands. Panel (12) plotted the microwave burst peak flux versus SXR peak flux in the two GOES bands. The cross and star symbols are microwave burst versus SXR flare of GOES 0.5-4{\AA} (A band) and 1-8{\AA} (B band), respectively. The correlation coefficients (ccA and ccB) between microwave burst and SXR flare in the two GOES bands are indicated. Red and black symbols distinguish the two event classes as in Figure.5.}
   \label{Fig6}
   \end{figure}

The microwave burst peak flux, fluence, energy and SXR peak flux, fluence of all ten events were calculated with the method illustrated in section 2. Figure 6 shows the relationship between microwave burst and GOES SXR flare. Panel (1)-(10) plotted the relationship between NORP microwave burst and GOES SXR flare, only seven events with NORP observation were including, while the other three events with only RSTN observation were not because the frequencies are different. The correlation coefficients between microwave burst fluence and SXR fluence are $>0.89$ in the frequency range of 2-35 GHz. While the correlation coefficients between microwave burst peak flux at single frequency and SXR peak flux are very poor or negative. The highest correlation coefficient is microwave fluence at 17 GHz and SXR fluence. Panel (11) and (12) shows that the correlation coefficients between microwave burst energy and SXR fluence are much higher than the correlation coefficients between microwave burst peak flux and SXR peak flux. The correlation coefficient between the microwave burst and the SXR flare in two bands are approximative.

\subsection{The relationships between microwave peak flux, fluence, energy and HXR peak flux, fluence}
\begin{figure}[ht]
   \centering
  \includegraphics[width=14.5cm, angle=0]{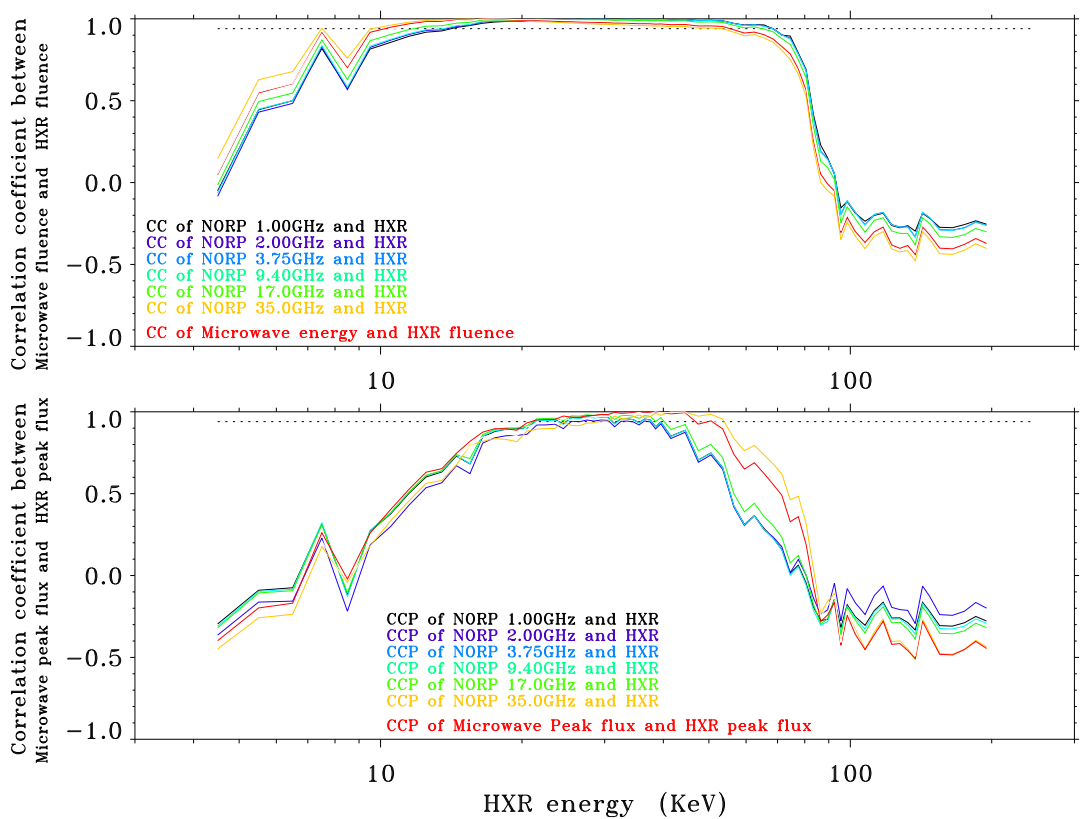}
   \caption{Upper panel plotted the correlation coefficients between microwave burst fluence at NORP 6 frequencies (different color) and HXR fluence in 76 raw bands within 4-250 keV. The red line is the correlation coefficients between microwave burst energy and HXR fluence. The dashed line marks the value of 0.94 which decided by all the cases within 15-50 keV. Bottom panel plotted the peak flux correlation coefficients between microwave burst at NORP 6 frequencies and HXR flare in 76 raw bands within 4-250 keV. The red line is the correlation coefficients between microwave burst peak flux (maximum value within NORP 1-35 GHz) and HXR peak flux.}
   \label{Fig7}
   \end{figure}

The HXR photons flux and fluence were calculated with solar software (SSW). Figure.7 plotted the correlation coefficients between microwave burst fluence or peak flux at NORP 6 frequencies and HXR fluence or peak flux in 76 raw bands within 4-250 keV. Only seven events with NORP observation were including. From the figure we can obtain results: 1) The correlation coefficients within 15-50 keV are higher than that $<$15 keV or $>$50 keV. The correlation below 10 keV or up 80 keV are poor or negative. 2) Below 80 keV, the fluence correlation coefficients is a little higher than peak flux correlation coefficients for most energy bands. 3) The peak flux or fluence correlation coefficients have no significant difference among different frequencies in the range of 1-35 GHz. However, in the HXR 40-80 keV band, there is a weak trend that peak flux correlation between microwave at 35 GHz and HXR are higher than that at low frequencies. 4) The correlation between microwave burst total energy and HXR fluence has no significant difference than the correlation between microwave burst fluence at 6 frequencies and HXR fluence. Also, the correlation between microwave burst peak flux (maximum value within NORP 1-35 GHz) and HXR peak flux has no significant difference than the correlation between microwave burst peak flux at 6 frequencies and HXR peak flux.

\section{Conclusion and Discussion}
\label{sect:discussion}
From the above multi-wavelength observations and statistical analysis on the energy and spectral characteristics of the confined flares we obtained the following conclusions:

(1) All the confined flare events were associated with microwave continuum burst extending to frequencies of greater than 9.4 GHz. Five events were associated with spectral fine structures at decimeter or centimeter wavelengths. With the exception of one single event where a type III burst was observed, the radio emission of the confined flares did not extend to meter wavelengths or beyond.

(2) The peak frequencies of all confined flare events are higher than 4.995 GHz and lower than or equal to 17 GHz.

(3) The radio burst fluence in the range 2-35 GHz of the events with spectral fine structures tends to be higher than that of the events without spectral fine structure. The trend is less clear for the peak flux.

As we mentioned above, it is interesting that there is one event accompanied by an IP type III burst at meter-to-kilometer waves. While about half of the flares of class M or above are accompanied by IP type III bursts (\citealt{Kou+2020}). Confined flares usually have no radio emission at meter wavelengths or beyond (\citealt{Gopa+2009,Klein+2010}), because the energetic electrons do not get access to magnetic structures that extend high into the corona. Nine out of the ten events studied here confirm this. The event with a type III burst was exceptional in that it was located in the periphery of the active region, rather than in the core (\citealt{Chen+2015}), i.e. at a place where energetic particles are indeed expected to get access to open magnetic field lines (\citealt{Axisa+1974,Zlob+1990,Poqu+McIn+1995,Hofm+Ruzd+2007}). The event is hence in line with the nine confined events, and confirms the picture of electron confinement when the energy release occurs in the core of strong magnetic fields.

It is not a surprise that half of the confined flares are accompanied by spectral fine structure, given that \citealt{Tanc+2019} reported some small or micro flares can still generate radio fine structure. The peak frequency of all confined events is  between $4.995$ and about $17$ GHz, and $\geq9$ GHz for nine of them. This is higher than the usual median range of $4.9\sim9$ GHz (\citealt{Nita+2004}). The higher peak frequency indicates stronger magnetic field in the source region, considering microwave bursts to be generated by the gyrosynchrotron mechanism (\citealt{Dulk+1985}). This indicates that the magnetic field surrounding the energy release region of a confined flare is stronger than that of the normal events. This result is consistent with earlier work on confined flares (\citealt{ Wang+Zhang+2007,Masson+2017,Amari+2018,Baum+2018,Li+2019}).
The radio burst energy of the events with fine structure tends to be higher than that of the events without fine structure. Such trend is more pronounced for the energy than that for the peak flux. We suggest that the occurrence probability of fine structure is more strongly related with the emitted microwave energy than with peak flux. The radio spectral fine structures are always generated by energetic electrons accelerated in the flare, and the radio spectral fine structure can be regarded as a signal of accelerated electrons in the flare event. Therefore the flare event with radio spectral fine structures may involve much more energy than that without radio spectral fine structures.

We analyzed the relationship between microwave burst and GOES SXR, the relationship between microwave burst and HXR, and found that:

(1) The correlation coefficients between microwave burst fluence (or total energy) and GOES SXR fluence are much higher than the correlation coefficient between microwave burst peak flux and GOES SXR peak flux.

(2) The correlation between microwave burst and HXR shows that: 1) The correlation coefficients within HXR 15-50 keV are higher than other energy bands. 2) For most energy bands below 80 keV, the correlation coefficients between microwave burst fluence (or total energy) and HXR fluence are a little higher than peak flux correlation coefficients. 3) In the HXR 40-80 keV band, there is a weak trend that peak flux correlation between microwave at 35 GHz and HXR are higher than that at low frequencies.

The microwave burst fluence or total energy is a much more precise estimate of the importance of an event than the peak flux. For example, the 2014 Oct.22 MF9 event is a moderate M2.7 flare with a very strong microwave burst with peak flux  $>2300$ sfu at 17 GHz. No radio emission is observed below 4 GHz. Figure.5 and figure.6 shows that the peak flux is high, while the fluence and total energy is low. Figure.7 shows high correlation coefficient ($\geq0.94$) between microwave burst fluence (or total energy) and HXR fluence in 15-50 keV. And they are a little higher than peak flux correlation between microwave burst and HXR flare for most energy bands below 80 keV. The peak flux correlation only shows high coefficient ($\geq0.94$) between microwave burst at 17-35 GHz and HXR flare in the 30-40 keV. This is a clue that microwave burst fluence or total energy can estimate a flare in a wide range globally, while peak flux can diagnose a flare partially.

In figure.7, the correlation coefficient between microwave burst fluence (or peak flux) and HXR fluence (or peak flux) is higher in the 15-50 keV range than 4-15 keV. This is consistent with the fact that both the higher-energy HXR emission and the microwave emission are produced by nonthermal electrons. The decreased correlation coefficient in the 50-100 keV range, and the negative coefficient above 100 keV indicates the lack of HXR photons with energy above around 50 keV in the considered confined flares. Figure.4 also shows that most of the confined flares had no significant HXR photons flux increase above 50 keV, that is, the background spectrum dominates above 50 keV. While \citealt{Kruc+2020} demonstrated the microwave (17 GHz and 34 GHz) peak flux shows good linear correlation with the nonthermal hard X-ray bremsstrahlung peak emission above 50 keV. Our finding of a reduced efficiency of electron acceleration to very high energies extends the finding by \citealt{Thal+2015} for the X1.6 flare on Oct.26.

In conclusion, the microwave burst energy (or fluence) and peak frequency are better than radio peak flux to estimate the power of flare. It is consistent with the result of \citealt{Grech+2015}. Thus in general, the microwave burst energy (or fluence) is one of the most important parameters determining ability of solar flare events to produce high energy particles. The energies to which electrons are accelerated in confined flares tend to be lower than in eruptive flares, while the magnetic field around the energy release region is stronger. However, only ten confined flares were studied in this work, and they belong to the same active region. In future work, we will extend the analysis to more confined flares, and other populations of flare events in more active regions.

\normalem
\begin{acknowledgements}
This work is supported by the NSFC grants 11790301, 11973057, 11941003, 11790305, 61811530282, Chinese-French cooperation between CNRS and NSFC, the MOST grant 2014FY120300, the National Key R\&D Program of China 2018YFA0404602, the International Partnership Program Of Chinese Academy of Sciences (Grant No.183311KYSB20200003), the Application and Foundation Project of Yunnan Province (Grant No.202001BB050032), the Commission for Collaborating Research Program of CAS KLSA, NAOC (Grant No.KLSA202115).
This work is also supported by ISSI-BJ, and partly supported by the international joint research program of the Institute for Space-Earth Environmental Research at Nagoya University and JSPS KAKENHI, grant No. JP18H01253. C. Tan thanks Dr. Lu, Lei for the helping of data analysis on Fermi/GBM.

\end{acknowledgements}

\bibliographystyle{raa}

\end{document}